\documentclass{moriond}

\usepackage{amssymb}

\def\be{\begin{equation}}
\def\ee{\end{equation}}
\def\bea{\begin{eqnarray}}
\def\eea{\end{eqnarray}}

\newcommand{\Photo}{}

\begin{document}
\vspace*{0cm}
\title{A TALE OF TWO SCALES: SCREENING IN LARGE SCALE STRUCTURE}

\author{MATTEO FASIELLO$^{a,c}$, ZVONIMIR VLAH$^{b,c}$}

\address{$^a$Institute of Cosmology \& Gravitation, University of Portsmouth, Portsmouth, PO1 3FX, UK\\  
$^b$Theoretical Physics Department, CERN, CH-1211 Geneve 23, Switzerland\\ $^c$Stanford Institute for Theoretical Physics and Department of Physics, Stanford University, CA, 94306}

\maketitle\abstracts{The perturbative treatment of dark matter in structure formation relies on the existence of a well-defined expansion parameter, $k/k_{\rm NL}$, with $k_{\rm NL}$ signalling the onset and ultimately the leading role of  non-linearities in the system. Cosmologies beyond the $\Lambda$CDM model often come with additional degree(s) of freedom. The scale $k_{\rm V}$ at which non-linearities become important in this additional sector(s) can be rather different from $k_{\rm NL}$.
For theories endowed with a Vainshtein-type screening mechanism, $k_{\rm V}$ sets the scale where screening becomes efficient and restores continuity with the predictions of general relativity. This is precisely the dynamics that allows such theories to pass existing observational tests at scales where general relativity has been tested with exquisite precision (e.g. solar system scales). We consider here the mildly-non-linear scales of a dark matter component coupled to a galileon-type field and focus in particular on the case of a $k_{\rm V}< k_{\rm NL}$ hierarchy.  We put forward  a phenomenological framework that describes the effects of screening dynamics on large scale structure observables.}

\centerline{ \textit{ $^*$Manuscript submitted to the Proceedings of the 43rd ``Rencontres de Moriond".}}

\section{Introduction}
The discovery in 1998 \cite{Perlmutter:1998np,Riess:1998cb} that the Universe is undergoing a period of accelerated expansion has been a momentous one for modern cosmology. General relativity with the help of a simple cosmological constant (c.c.), $\,\Lambda$,  is able to account for such acceleration. However, the value $\Lambda_{\rm obs}$ inferred from observations is at odds with the expectation that any contribution to the vacuum energy (e.g. from Standard Model particles) would also automatically contribute to the cosmological constant, setting a lower bound for the c.c. which is many orders of magnitude above $\Lambda_{\rm obs}$. This giant mismatch is know as the c.c. problem \cite{Weinberg:1988cp}. Several ideas have been put forward to make sense of late-time acceleration: from the presence of extra scalar field(s) driving acceleration (much like for inflationary models), to infrared modifications of gravity, to an anthropic perspective on the value of $\Lambda_{\rm obs}$, the latter being best understood within the string theory landscape program \cite{Susskind:2003kw}.\\
In what follows we will focus  on infrared modifications of gravity. These often come with extra degrees of freedom driving the acceleration. The key point is that gravity is well-tested in high-density, small-scale, environments such as the solar system, but there are far less constraints at cosmological scales \cite{Koyama:2015vza}. So long as a mechanism (i.e. screening) is in place that allows significant extra dynamics at large scales and suppresses it at smaller ones, gravity is modified only in the infrared and the corresponding model can be viable.

\section{The quasi-linear regime of structure formation and beyond-$\Lambda$CDM physics}
The dynamics of the extra fields in beyond-$\Lambda$CDM theories is then expected to be most evident at large scales and to impact structure formation. For sufficiently small scales, screening will restore LSS to its $\Lambda$CDM dynamics. In Vainshtein theories \cite{Vainshtein:1972sx}, screening is an intrinsically non-linear phenomenon and becomes efficient when non-linearities in the ``Vainshtein" sector become relevant. Let us illustrate this case in the specific example  of a dark matter (DM) component coupled to a cubic galileon field \footnote{Galileons naturally emerge in theories such as massive gravity \cite{deRham:2010kj} and DGP \cite{Dvali:2000hr}. They provide the most natural realization of Vainshtein/kinetic screening.}:  
\bea
\frac{\partial \delta_m}{\partial \tau}+\partial_i [(1+\delta_m) v_m^i]=0\; ; \qquad \frac{\partial v_m^i}{\partial \tau}+ \mathcal{H}v_m^i+v_m^j \partial_j v_m^i  =-\nabla^i  \Phi
\label{1}
\eea
\bea 
\nabla^2 \Phi= \frac{3}{2}\mathcal{H}^2\Omega_m \delta_m +F(\bar{\phi}) \nabla^2\delta\phi\; ; \qquad \nabla^2 \delta\phi +{\rm non\,linearities}= \frac{\beta}{M_{\rm Pl}}\,\delta_m \label{2}
\label{2}
\eea
The relations in Eq.~(\ref{1}) are respectively the continuity and Euler equations for DM. The first term on the right hand side of the Poisson equation would normally close the system. The coupling with a galileon-like field $\phi$ modifies the Poisson equation providing a contribution that mimics an additional gravitational potential, hence we see how this and similar theories support a fifth force \footnote{Already a minimally coupled scalar, such as in quintessence models, can modify the dynamics at large scales$\;$\cite{Fasiello:2016qpn,Sefusatti:2011cm,Anselmi:2011ef,Fasiello:2016yvr}. However, the difference between the linear and highly non-linear regimes is less striking in such cases.}. The last relation in Eq.~(\ref{2}) is the galileon own equation of motion. It is clear that, the parameter $\beta$ permitting, the linear solution for the galileon field will generally lead to an order one modification of the Poisson equation and consequently visibly impact the dynamics of matter.\\ The effect of non-linearities in the Vainshtein (i.e. galileon) sector is to suppress the impact of $\phi$ on the Poisson equation and therefore on DM. Under certain simplifying assumptions, exact solutions exist for this system \cite{Koyama:2007ih}: within a certain radius from the source, known as Vainshtein radius $r_{\rm V}$, the fifth source is suppressed and $\Lambda$CDM behaviour is recovered; for $r> r_{\rm V}$ there are order one differences from standard DM dynamics and structure formation. The scale $k_{\rm V}$ is the natural counterpart of Vainshtein radius in Fourier space. Let us inspect this quantity more closely by writing explicitly the cubic galileon equation of motion:
\bea
\nabla^2 \phi+ \underbrace{\frac{1}{\Lambda^3} \Big[ (\nabla^2 \phi)^2 - (\nabla_i \nabla_j \phi)^2\Big]}_{\rm n-l}=\beta \frac{\rho_m}{M_{\rm Pl}} \; .
\label{3}
\eea
It is clear that the momentum $k_{\rm V}$ above which the non-linear piece becomes important will depend on: (i) the ``universal" quantity $\Lambda$, inherited from the Lagrangian \footnote{In the case of massive gravity, for example, one finds $\Lambda\equiv\Lambda_3=(m^2 M_{\rm Pl})^{1/3}$, with $m$ the bare graviton mass.} and not to be confused with the cosmological constant; (ii) the background time-evolution of the field $\phi$ and (iii) the source, just as is the case for $r_{\rm V}$.
\noindent The asymptotic $k$-behaviour of our system in Eqs.~(\ref{1},\ref{2}) is then clear: -- for $k\ll k_{\rm V}$, recalling that $k_{\rm V}< k_{\rm NL}$, perturbation theory (PT) is consistent both in $k/k_{\rm V}$ and $k/k_{\rm NL}$ and one expects $\mathcal{O}(1)$ modifications to DM dynamics; -- for $k \gtrsim 
k_{\rm V}$ instead, $\Lambda$CDM is recovered.\\ In the absence of exact solutions, how do we describe the transition between these two regimes?$\,$\footnote{Note that, crucially, the unscreened $\leftrightarrow$ screened transition can occur already at quasi-linear scales in the sense of the $k/k_{\rm NL}$ expansion.} Clearly, given an order one expansion parameter $k/k_{\rm V}\lesssim 1$,  an exclusively-perturbative approach is not appropriate. We will then \textit{resum} the expansion in $k/k_{\rm V}$. In order to do so, we will focus on a specific observable such as the power spectrum of the total density contrast and assume a \textit{perturbative} solution (in $k/k_{\rm V}, k/k_{\rm NL}$ ), namely $P^{(n)}_{\rm pert}$ , is available up to a given order. Formally, the resummation in $k/k_{\rm V}$ can be written as
\bea
P_{\rm res} \big |_N (k,\tau)= \sum_{n=0}^N P_{\rm res}^{(n)} (k,\tau) = \sum_{n=0}^N \int \frac{d^3 k'}{(2\pi)^3} ~\mathcal K^{N}_n(k',k,\tau) P^{(n)}_{\rm pert}(k^{\prime},\tau) ,
\label{4}
\eea
where we have introduced the kernels $\mathcal K^{N}_n$ to account for the resummation. The reader might recall that a similar formulation has been used in the context of the so-called baryonic acoustic oscillations (BAO) resummation scheme \cite{Senatore:2014via,Vlah:2015sea,Blas:2016sfa}. The physics under scrutiny here is of course rather different from BAO dynamics but the key idea, namely the existence of an $\mathcal{O}(1)$ parameter whose PT needs to be resummed, is the same. In our setup we find it convenient to simplify Eq.~(\ref{4}) to 
\bea
P_{\rm res} \big |_N (k,\tau) = \sum_{n=0}^N \Big[  P_{\Lambda\rm{CDM}}^{(n)} (k,\tau) + K^N_{n}(k,\tau) \Delta P^{(n)} (k,\tau) \Big],
\label{5}
\eea
where $\Delta P^{(n)}\equiv P^{(n)}_{\rm pert}- P^{(n)}_{\rm \Lambda CDM}$. The asymptotic behaviour of our coupled system, together with the symmetries of the problem, suggest the following form for the kernels:
\bea
K_{\rm Gau.}(k,\tau) = \exp \left(- \sum_m \alpha_m(\tau) (k/k_{\rm V})^{2m} \right)\; ;\;
K_{\rm Lor.}(k,\tau)= 1/\left(1 + \sum_m \alpha_m(\tau) (k/k_{\rm V})^{2m} \right)\, .
\label{6}
\eea
\noindent We pause here to note that the phenomenological nature of our approach is underscored by the following facts: (i) we are not deriving the kernels from first principles but rather putting them forward on the basis of asymptotic behaviour and symmetries; (ii) we are choosing a constant $k_{\rm V}$ and accounting for its time dependence through the coefficients $\alpha_m(\tau)$; (iii) in an exact derivation one would expect the kernels to be applied directly at the level of the fields rather than on observables such as the power spectrum (the latter would be proportional to a \textit{convolution} of kernels).

\begin{figure*}[h!]
\centerline{\includegraphics[scale=0.6]{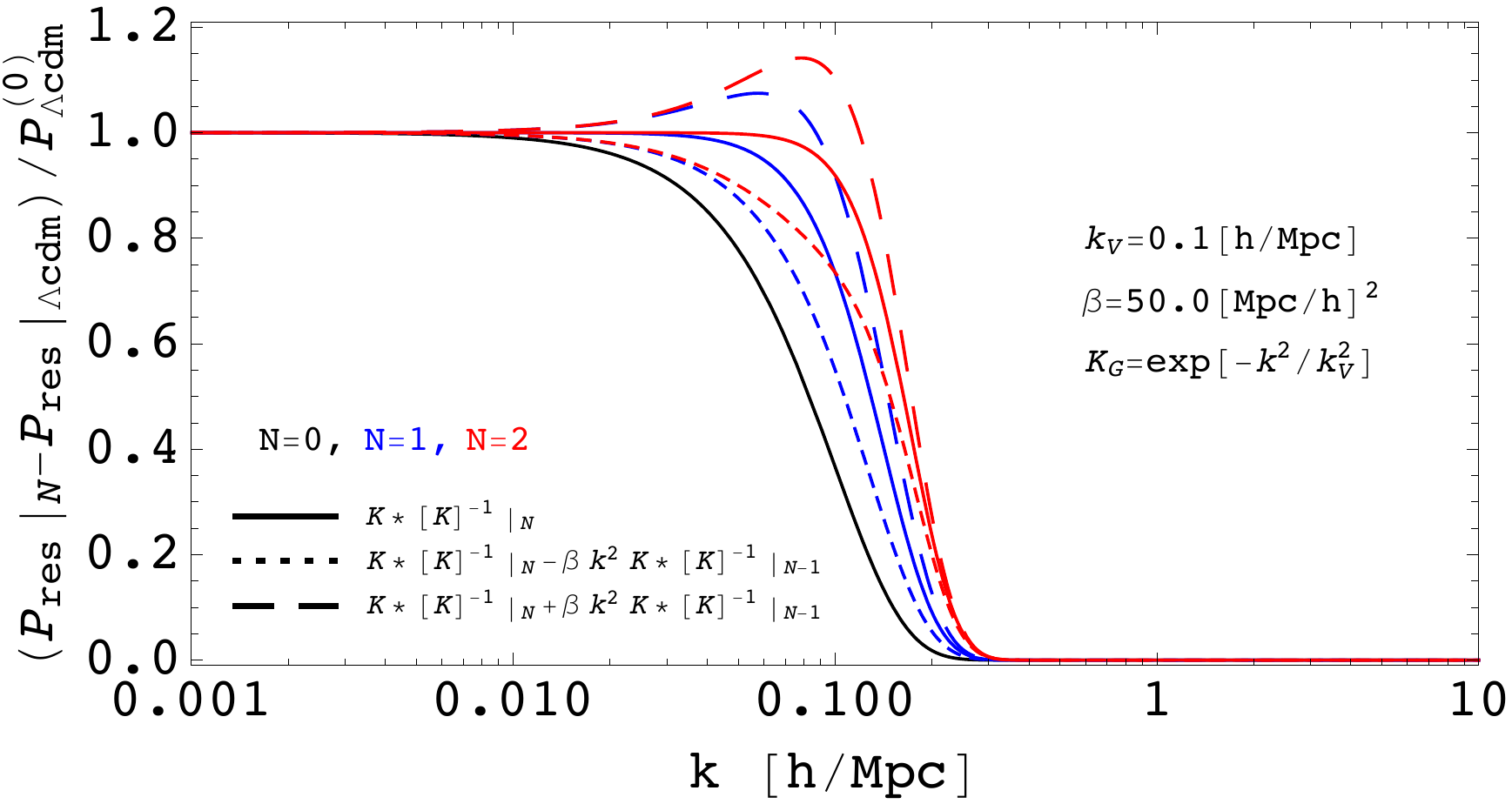}{\centering}}
\caption{The fractional difference is plotted between a fully screened total power spectrum and the $\Lambda$CDM case. Different colors indicate here the different perturbative orders. The action of the Gaussian kernel is highlighted by continuous, dotted, and dashed lines. In this approximation $K_{\rm G}$ is regulated by the value of just one parameter.}
\label{fig}
\end{figure*}

Let us now add a rather technical ingredient to our framework:
\bea
K^N_n(k,\tau) = K (k,\tau) \big[ K \big]^{-1}\Big|_{N-n}  (k,\tau),
\label{7}
\eea
with the last term being the $(N-n)-$th order Taylor expansion of the inverse of the reduced kernel $K^N$, function of time and  $k/k_{\rm V}$. Eq.~(\ref{7}) outlines the need for both an ``N" and ``n" indices in the kernels:  $N$ stands for the PT order we are working at and  $n$ refers to the expansion in $k/k_{\rm V}$: as we go higher in PT (higher $N$), more of the screening is captured already perturbatively and so the $N$-dependence in the kernels ensures that no ``double counting"  (perturbative \textit{plus} re-summed) of screening contributions takes place. The same line of reasoning leads one to favour the Gaussian kernel over the Lorentzian one in Eq.~(\ref{6}).\\ Let us specialize to Gaussian kernels and spell out the action of our screening ``filter" on observables such as the power spectrum:
\bea
P_{\rm res} \big |_N (k,\tau)= P_{\Lambda\rm{CDM}} \big |_N (k,\tau) \hspace{0cm} +  K (k,\tau) \sum_{n=0}^N \big[ K \big]^{-1}\Big|_{N-n}  (k,\tau) \Delta P^{(n)} (k,\tau)
\label{8}
\eea
Under extra assumptions \footnote{Specifically, the following relations are assumed to hold: $\Delta P^{(0)} \sim\,P^{(0)}_{\rm \Lambda CDM}$ and $\Delta P^{(n)} \sim  k^2 \Delta P^{(0)} \sim k^2  P^{(0)}_{\rm \Lambda CDM}$. See \cite{Fasiello:2017bot} for further details on this point.} we are able to illustrate, by means of Fig.~(\ref{fig}) how the fractional difference between a fully-screened total power spectrum and the $\Lambda$CDM power spectrum would

\noindent look like when Vainshtein screening takes place at $k_{\rm V}=0.1[{\rm h/Mpc}]$, a regime where the $k/k_{\rm NL}$ expansion is still well under perturbative control. Comparison with simulations \cite{Falck:2014jwa} of Vainshtein screened theories such as DGP are very encouraging as to the effectiveness of the framework outlined here.



\section*{Acknowledgments}

It is a pleasure to thank Diego Blas and Martin White for illuminating comments on related work.

\end{document}